# Bottom-up fabrication of atomically precise graphene nanoribbons.

**Martina Corso, Eduard Carbonell-Sanromà and Dimas G. de Oteyza**

**Abstract** Graphene nanoribbons (GNRs) make up an extremely interesting class of materials. On the one hand GNRs share many of the superlative properties of graphene, while on the other hand they display an exceptional degree of tunability of their optoelectronic properties. The presence or absence of correlated low-dimensional magnetism, or of a widely tunable band gap, is determined by the boundary conditions imposed by the width, crystallographic symmetry and edge structure of the nanoribbons. In combination with additional controllable parameters like the presence of heteroatoms, tailored strain, or the formation of heterostructures, the possibilities to shape the electronic properties of GNRs according to our needs are fantastic. However, to really benefit from that tunability and harness the opportunities offered by GNRs, atomic precision is strictly required in their synthesis. This can be achieved through an on-surface synthesis approach, in which one lets appropriately designed precursor molecules to react in a selective way that ends up forming GNRs. In this chapter we review the structure-property relations inherent to GNRs, the synthesis approach and the ways in which the varied properties of the resulting ribbons have been probed, finalizing with selected examples of demonstrated GNR applications.

## 1 Introduction

Graphene has attracted enormous interest since its first experimental realization through exfoliation of graphite, mainly because of its many superlative properties [1,2]. By way of example, it is the thinnest, lightest and strongest material known.

M. Corso • D. G. de Oteyza
Centro de Física de Materiales CSIC-UPV/EHU – Materials Physics Center, Paseo Manuel Lardizabal 5, 20018 San Sebastián, Spain
e-mail: martina.corso@ehu.eus, d_g_oteyza@ehu.es

M. Corso • E. Carbonell-Sanromà
CIC nanoGUNE, Avenida Tolosa 76, 20018 San Sebastián, Spain

D.G. de Oteyza
Donostia International Physics Center, Paseo Manuel Lardizabal, 4, 20018 San Sebastián, Spain

D.G. de Oteyza
Ikerbasque, Basque Foundation for Science, 48011 Bilbao, Spain

It is also the material with highest thermal conductivity and with highest electron mobility. The latter causes graphene to be considered as a potentially revolutionary material for future electronic applications [2,3]. In this respect, however, graphene also faces a drawback for its implementation in conventional electronics: the lack of a band gap to turn the electron conduction on or off [3].

Various strategies have been investigated to overcome this limitation, and one of them is through electron confinement in nanostructured graphene. In particular, graphene nanoribbons, narrow 1D stripes of graphene with widths in the nanometer range, display remarkably varied electronic properties depending on their particular structure at the very atomic level [4].

A visual framework to understand this is analyzing the aromaticity of some representative GNR structures [5,6]. According to Kekulé, each carbon in polycyclic aromatic hydrocarbons (PAH) like GNRs or graphene itself has its four valence electrons arranged in single, double or triple bonds with electrons from neighboring atoms. The structure of a PAH is then the superposition of all possible Kekulé bond configurations. Within this picture, the delocalization of six π-electrons in a carbon hexagon due to the resonance of two Kekulé configurations with alternating single and double bonds is called a Clar sextet, pictured as a circle within the corresponding hexagon (Fig. 1a). According to Clar´s theory [7], the most representative and stable structure of a PAH is that with the highest number of Clar sextets, which is called the "*Clar formula*". Here it is important to keep in mind that the bonds sticking out of a Clar sextet are formally single bonds and thereby impede two neighboring hexagons to be Clar sextets simultaneously.

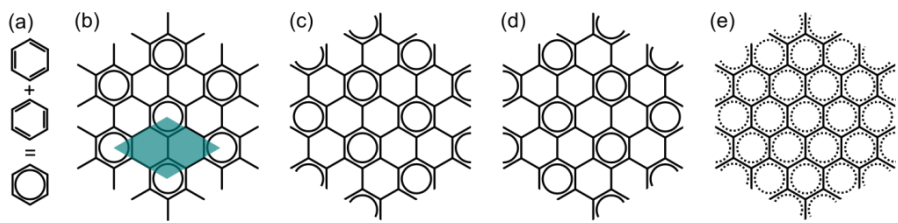

**Fig. 1 a** Graphical explanation of the Clar sextet as being the superposition of two resonant Kekulé configurations. **b-d** Each of the three equivalent Clar formulas of graphene, with its √3×√3 superstructure highlighted in cyan in b. **e** Superposition of all three Clar formulas. Dotted circles correspond to Clar sextets within graphene, although not all simultaneously.

Graphene can be represented with three equivalent Clar formulas, in each of which one out of every three hexagons is a Clar sextet, arranged in a (√3×√3)R30º superstructure (Fig. 1b-d). Considering a combination of the three possible Clar formulas, all hexagons in graphene can be Clar sextets and are fully equivalent (Fig. 1e). In addition, bonds within a Clar sextet are equivalent as well. Since bond length alternation (BLA) is a measure of the aromaticity of a system and one of the main causes for the opening of band gaps in conjugated organic materials [8,9], graphene is an excellent example of a perfectly aromatic system with absent BLA and zero band gap.

Now we place our focus on GNRs. Because GNRs include many different structures, a first classification of GNRs is typically performed depending on their edge orientation. They are called zigzag nanoribbons (zGNR) when the edge runs parallel to one of graphene´s lattice vectors, or armchair nanoribbons (aGNR) when the edges run along the high symmetry direction thirty degrees off the lattice vectors (Fig. 2a). GNRs with edges running along any intermediate orientation are called chiral nanoribbons. Their chirality is characterized either by the angle of the edge orientation with respect to a graphene lattice direction, or by the edge unit cell vector in terms of graphene lattice vectors (Fig. 2a). The key in GNRs is that the presence of edges imposes boundary conditions which, among other effects, limit the number of possible Clar sextets.

The width of aGNRs is most commonly given as the number of dimer lines $N_a$ present across the ribbon (Fig. 2b). In the following we assume the most common scenario of purely $sp^2$ hybridized carbon atoms within the ribbons, consequently with singly hydrogenated edges. Under these conditions, we will see how the boundary conditions divide the aGNRs into three families, depending on the number of Clar formulas they display [5,6].

For aGNR with $N_a = 3p$, $p$ being an integer, there is only one possible Clar formula maximizing the number of Clar sextets. In this formula, all π-electrons participate in the Clar sextets, consequently displaying no localized double bonds (Fig. 2b). While the π-electrons are delocalized within the Clar sextets, the bonds sticking out of them remain single bonds, causing a substantial variation of the mean bond length within the ribbons that opens a considerable band gap [5,6]. For aGNRs with $N_a = 3p+1$ there are two possible Clar formulas, both of which include localized double bonds (Fig. 2b). A combination of the two Clar formulas still leaves hexagons with and without Clar sextets, causing again notable bond length variations within the ribbon. This bond length variation and the electron localization in the structure´s double bonds cause this family to have an even larger band gap than the $3p$ family [5,6]. The scenario is very different with aGNRs of the $N_a = 3p+2$ family. In this case, there are many different Clar formulas, each of them featuring two localized double bonds (Fig. 2b). However, a linear combination of all the Clar formulas ultimately renders a highly aromatic structure with little bond length variations along the middle part of the ribbon. This implies a high degree of electron delocalization and consequently a low band gap [5,6].

It is very interesting to analyze zGNRs in this same framework, since they can host only a limited amount of Clar sextets even for infinitely long ribbons (Fig. 2c). Because sextets can be placed anywhere along the longitudinal axis of the ribbon, there are infinite Clar formulas for zGNRs. Their combination ends up being equivalent to a superposition of two fully quinoidal structures, that is, structures with two double bonds per carbon hexagon (Fig. 2c). However, allowing the introduction of unpaired electrons (radicals) into the structure, Clar sextets can be distributed anew along the whole ribbon length (Fig. 2c).

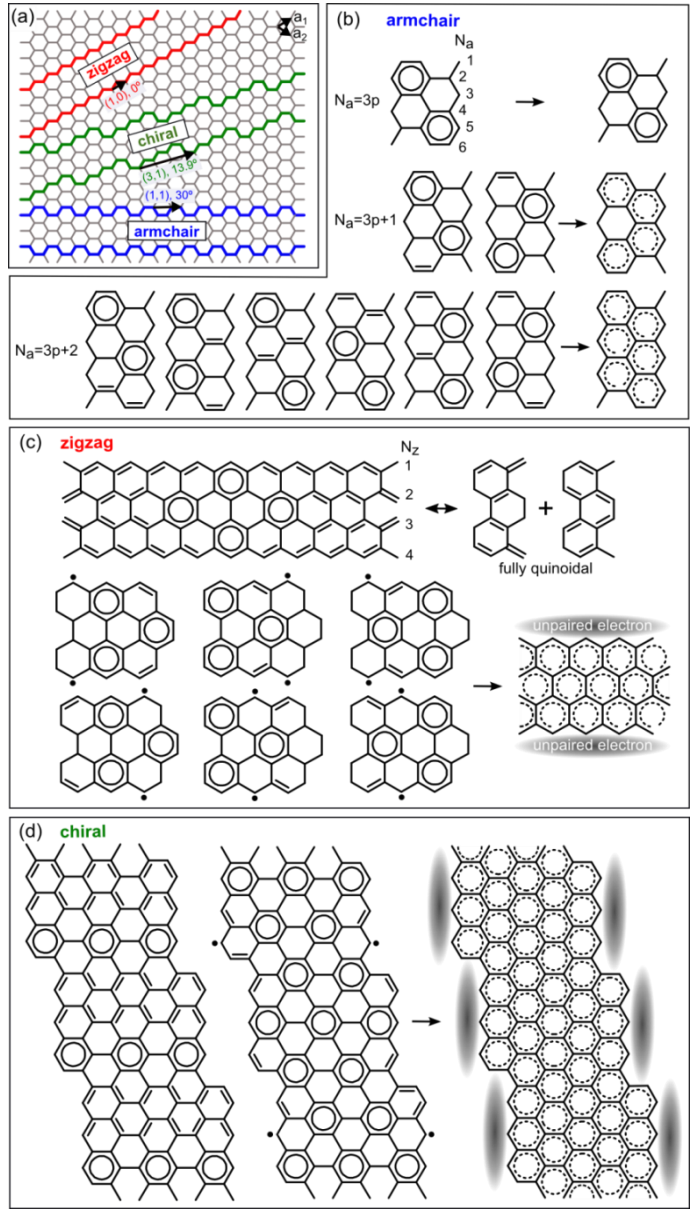

**Fig. 2 a** Clasification of GNRs by edge orientation into zigzag, chiral or armchair. The respective edge periodicity is given by the unit cell vector in terms of graphene´s lattice vectors $a_1$ and $a_2$. **b** Clar sexted distribution on aGNR, classified into three families according to its width in terms of dimer lines $N_a$. The family with $N_a=3p$ possess only one formula, exemplified in 6-aGNRs. The family with $N_a=3p + 1$ possess two compatible Clar formulas. In each of these formulas there is a localized double bond at the edge of the ribbon, as exemplified with the formulas of 7-aGNRs. On the right, a superposition of both Clar formulas is given, showing all hexagons hosting a Clar sextet (although not all simultaneously). The family with $N_a=3p + 2$ ribbons have more than two Clar formulas, namely 3p. Some of them are shown for 8-aGNRs. On the right, a superposition of all Clar formulas shows the Clar sextets delocalized along all carbon hexagons. **c** Clar sextet distribution on zGNRs. As exemplified with 4-zGNRs, only a limited number of Clar sextets exists even on infinitely long ribbons. Infinite Clar formulas exist, due to their arbitrary location along the ribbon, and the superposition is equivalent to the two quinoidal structures drawn. Below it is shown how by adding unpaired electrons, Clar sextets can be redistributed along the whole ribbon. Some of the Clar formulas are shown, as well as their superposition evidencing delocalized Clar sextets along the ribbon and unpaired electrons along the edges. **d** Clar sextet distribution on chiral GNRs, evidencing a limited number of Clar sextets per zigzag segment and how addition of unpaired electrons allows creation of additional Clar sextets. One of the multiple Clar formulas is shown for each case, including on the right a superposition of all the possible Clar formulas that evidences the delocalized Clar sextets along the ribbon and unpaired electrons along its edges.

This configuration reduces the overall ribbon energy and readily provides an explanation of some of most intriguing properties of zGNRs. A combination of all possible Clar formulas evidences all hexagons as being Clar sextets, implying a high degree of aromaticity and electron delocalization along zGNRs, as well as an associated low band gap. In addition, because the radicals are always located on the edges, combination of the various Clar formulas evidences unpaired electrons fully delocalized along both ribbon edges. This is the origin of the edge states in zGNR, appearing, as expected from unpaired electrons, at the Fermi level. Finally, since unpaired electrons translate to an uncompensated spin, Clar´s formalism also implies the spin polarized nature of the edge states in zGNRs.

Last but not least, chiral GNRs display a combination of alternating zigzag and armchair segments along their edges. The lower the chiral angle the longer the zigzag segments are, and viceversa. As pictured on the (3,1) chiral nanoribbon example in Fig. 2d, there is a limited number of Clar sextets within each zigzag segment. However, as occurs for pure zGNRs, addition of unpaired electrons on the zigzag edges allows formation of additional Clar sextets (Fig. 2d). The combination of the different Clar formulas evidences the high degree of aromaticity and electron delocalization in chiral GNRs that explains the low band gap, as well as the creation of delocalized and spin polarized edge states along the zigzag segments of the ribbon (Fig. 2d). At this point it is important to remind the intricate dependence of these edge states with the detailed GNR structure. In infinitely long zGNRs, addition of unpaired electrons implies changing from a finite number of sextets to an infinite number of sextets, making the edge state creation always favorable. Instead, infinite chiral GNRs readily display infinite Clar sextets (with a limited number of them on each zigzag segment, as mentioned above). Addition of unpaired electrons simply increases the number of Clar sextets per zigzag segment, or in other words the density of Clar sextets per GNR length unit. This has the following implications. On the one hand, there is a width threshold below which the edge states are not present because for narrow ribbons the energy cost of having unpaired electrons may not be compensated by the creation of an insufficient number of new Clar sextets. On the other hand, the chiral angle also plays a major role. The lower the chiral angle (longer zigzag segments), the more new sextets can be created by addition of unpaired electrons, and therefore the lower the width threshold for edge state creation. Lastly, the lower the chiral angle the larger is also the number of unpaired electrons per edge atom and consequently also the net spin moment on each nanoribbon edge.

A similar analysis based on the Clar sextets and the resulting GNR´s aromaticity has been also applied to nanoribbons doped with heteroatoms [10]. Depending on the particular atomic species and its bonding site, the Clar sextet distribution and the associated aromaticity can change substantially, having a direct and notable impact on the GNR´s electronic properties [10].

All the above shows, from an intuitive and easy to visualize chemical viewpoint, one of the main virtues of GNRs: their amazing variety of electronic properties, ranging from large band gap semiconductors to gap-less structures with spin-polarized edge states. However, it also remarks the stringent need for atomic precision in their synthesis, since minute structural changes at the atomic level can cause unproportioned changes in their electronic properties.

## 2 Synthesis

Current digital logic devices require excellent switching capabilities and an on-off ratio ($I_{on}/I_{off}$, corresponding to the current flows under *on* and *off* operation conditions) in the order of $10^4$ to $10^7$. The latter in turn requires band gaps of 0.4 eV or more [3]. If graphene nanoribbons are to be integrated into such structures, their width needs to be scaled down to few nanometers [11]. This sets challenging boundary conditions to be solved in the synthesis of graphene nanoribbons, complicated even further by the requisite of atomic precision described in the previous section.

Several groups have readily reported the successful synthesis of GNRs by top-down methods and their subsequent characterization. Amongst the different strategies we find e.g. the unzipping of carbon nanotubes [12], etching of 2D-graphene [13], chemical vapor deposition [14], or scanning probe lithography [15]. However, in spite of the important insight into the properties of GNRs provided in these works, such techniques lack the synthetic reproducibility and/or the atomic precision, the latter being of particular importance to be able to aim at rationally chosen

GNR structures (e.g. atomically precise width control) and to avoid the often dominating disorder effects [16,17].

Instead, Cai and co-workers reported in a seminal work a bottom-up strategy to do just that: selectively grow atomically precise nanoribbons with widths in the nanometer range [18]. That inspiring work has sparked the interest and research efforts in this direction, which basically consists in letting appropriately designed precursor molecules to react in a selective way that ends up forming GNRs [4,19]. This approach has been followed in solution [19], as well as supported on solid surfaces [4,19]. Henceforth we will focus on the latter, for which no additional functionalization with side-chains is needed to provide solubility to the GNR products. In a first step, the reactants are deposited on a surface. Thereafter, they are externally activated, triggering polymerization reactions and their subsequent transformation into GNRs. Common to all substrate-supported GNRs successfully synthesized to date, the reactant deposition was by molecular beam deposition, the polymerization based on Ullmann coupling, and the reaction thermally activated. Nevertheless, we would like to remark that there may be useful alternatives to each of these, as are e.g. electrospray deposition [20], other C-C coupling reactions like C-H activation [21] or enediyne cyclizations [22], and photoactivation of the reactions [23,24]. Such alternatives may help reducing the current constraints on reactants and substrates.

A prototypical reaction process is for example that shown in Fig. 3a [25], which includes polymerization through Ullmann coupling, but also a second cyclo-dehydrogenation step. The whole reaction process has been reviewed extensively [4,21,26] and here we only provide a brief description of the main steps. Ullmann coupling itself readily involves various steps [26,27]. First there is a reactant dehalogenation, resulting in surface or adatom-stabilized radical intermediates. This process is catalyzed by metallic substrates [27], allowing the dehalogenation to occur at temperatures below the temperatures of molecular desorption or of other unwanted side-reactions that could compromise the synthetic selectivity and the overall polymerization process. In a next step, the radical intermediates diffuse along the surface to meet each other and bind covalently. In between the latter two, and depending on the type of substrate, the formation of a metastable metal-organic intermediate is often observed [21,26]. Finally, cyclo-dehydrogenation leads to the planarization of the polymers, ending up in atomically precise graphene nanoribbons with their structure univocally defined by the design of the reactant. Annealing at higher temperatures can also cause the coupling of neighboring GNRs by inter-ribbon cyclo-dehydrogenation [28-30]. This process, however, results in the fusion of a random number of GNRs, lacking selectivity for the formation of precise widths and generating substantial disorder.

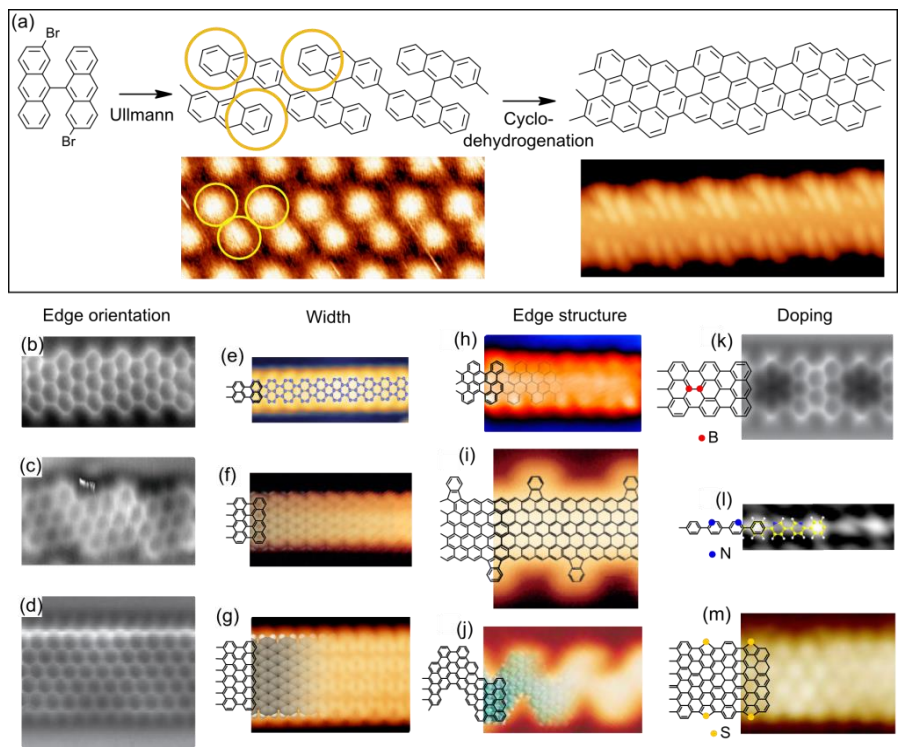

**Fig. 3 a** Schematic representation through wireframe diagrams of an archetypical reaction process. The precursors first undergo Ullmann coupling, resulting in non-planar polymeric structures due to intramolecular steric hindrance. The circles highlight sections that are elevated with respect to the surface. In a second step the planar GNR is formed through cyclodehydrogenation. Associated STM images revealing the non-planar and planar nature of polymer and GNR are added below the wireframe structures. The use of different precursors allows growing different GNRs with varying edge orientation (**b-d**), width (**e-g**), edge morphology (**h-j**) or doping (**k-m**). (a) Reprinted with permission from [25]. (b) Reprinted by permission from Macmillan Publishers Ltd: Nature Communications [40], copyright (2013), (c) Reprinted with permission from [41]. Copyright (2016) American Chemical Society. (d) and (i) Adapted by permission from Macmillan Publishers Ltd: Nature [42], copyright (2016). (e) Adapted with permission from [43]. (f) Reprinted with permission from [45]. Copyright (2017) American Chemical Society. (g) Reprinted with permission from [46]. Copyright (2013) American Chemical Society. (h) Reprinted with permission from [47]. (j) Adapted by permission from Macmillan Publishers Ltd: Nature [18], copyright (2010). (k) Adapted with permission from [48]. (l) Adapted with permission from [36]. (m) Adapted with permission from [49]. Copyright (2016) American Chemical Society.

Many of the processes described above and their associated energy barriers depend notably on the substrate material and surface structure, underlining their key role in the overall synthesis. While on the one hand it limits the range of potential substrates to be used, on the other hand their appropriate choice can be used as an additional parameter to control the reaction and thus the resulting products [31,32], their distribution or alignment [33-36].

Obviously, the particular reactant used also has a notable effect on the various chemical processes and their barriers. By way of example, one can compare the coupling of 10,10´-dibromo-9,9´-bianthracene [18] and the mono-anthracene equivalent. The latter lies flat on the surface, and the radical coupling is therefore hindered by the steric repulsions between the hydrogens surrounding the carbon radical. Instead, the intramolecular steric hindrance between its two anthracene units makes the former a non-planar precursor. As a result, the steric hindrance between two radical intermediates is reduced, allowing them to get close enough to form the covalent C-C bond [4].

Another beautiful example is the comparison of the same 10,10´-dibromo-9,9´-bianthracene designed to render 7-aGNRs, with 2,2′-dibromo-9,9′-bianthracene, designed to render chiral (3,1)-GNRs [25]. For the latter, the Br atoms lie closer to the surface, lowering the temperature threshold for dehalogenation. However, even more notable is the change in the temperature threshold for the cyclo-dehydrogenation. In this case it rather relates to the completely different strain between the two resulting polymers. With the former, the anthracene units are linked covalently along their short axis by a bond that allows free rotational movement with respect to their neighbors. This freedom results in alternatively tilted anthracene units along the polymer backbone so as to minimize the steric hindrance from opposing H atoms. Instead, with the latter the anthracene units are linked covalently to their neighbors both along their long and short axes (Fig. 3a). Thus, although the anthracene units still display the same alternative tilt to reduce the steric hindrance, the covalent bonds along the long anthracene's axes limit the structure's rotational freedom, resulting in a substantially strained geometry. Sterically induced strain is known to weaken the involved C-H bonds and thereby lower the cyclo-dehydrogenation barriers [37-39], in turn explaining the substantially lower threshold temperature for 2,2′-dibromo-9,9′-bianthracene [25].

Effects like these can be of great importance for potential applications. Focusing on the last example, the lower cyclo-dehydrogenation temperature may on the one hand allow the use of different substrates that could not stand higher temperatures. On the other hand, it may have a strong impact on the resulting products. By way of example, on Au(111) it brings the threshold temperatures for Ullmann coupling and cyclo-dehydrogenation close to each other. Under this scenario, radical quenching by liberated H atoms competes with the radical step growth polymerization, greatly lowering the average length of the resultant GNRs [25].

However, most importantly, the appropriate design of precursors allows growing GNRs with different and tailored structures, all of them with atomic precision. This is shown in Fig. 3, displaying selected GNRs with different edge orientations (Fig. 3b-d) [40-42], different width (Fig. 3e-g) [43-46], different edge structure (Fig. 3h-j) [18,42,47] and different doping (Fig. 3k-m) [36,48,49], all of them allowing to tune in a controlled way the electronic properties of GNRs.

# 3 Characterization tools and associated insight

## *3.1 Graphene nanoribbon´s electronic properties determination*

The electronic structure of GNRs and in particular their band gap ($E_g$) size and the values of bands' effective masses ($m_{VB}$ and $m_{CB}$), have been addressed locally with scanning tunneling spectroscopy (STS) [34,40,42,43,45,46,50,51], and with average experimental techniques as angle-resolved-, inverse- and two-photon-photoemission [34,35,52-54], high-resolution electron energy loss [52,55] and optical spectroscopy [56].

The most widespread method used so far is STS. It allows determining the electronic band dispersion of occupied and empty energy levels for the very same GNR. A precise estimation of bands position and effective masses is given by means of Fourier-transformed scanning tunneling spectroscopy (FT-STS) [57]. STS measures the local density of states (LDOS) as a function of tip position and applied sample bias. The presence of edges in GNRs gives rise to standing wave patterns that arise due to scattering of electronic Bloch wave functions against such termini. Recording standing waves for different energies along one GNRs edge, allows determining the *energy vs momentum* dispersion of GNRs' states via Fourier transform processing of STS data. Precise bandgap values of $E_g$=2.37±0.06 eV and $E_g$=1.38±0.03 eV have been extrapolated from valence band maximum and conduction band minimum for 7-aGNR (Fig. 4c-f) [50] and 9-aGNRs [45] grown on Au(111), respectively.

Such precise value could not be simply inferred by single STS spectra (Fig. 4a). On one hand, this is due to the fact that the onset of the GNR's electronic states could be ambiguous due to fingerprints of the electronic structure of the substrate. On the Au(111) surface, for example, the presence of the surface state at -0.45 eV makes it difficult to identify the valence band position in many GNRs. On the other hand the signal from GNRs' valence or conduction bands could be too weak to be detected by STS due to the symmetry of the GNRs' orbitals and the LDOS decay above the GNRs' plane [45,50].

The occupied frontier band of GNRs have been measured in several cases by angle resolved photoemission (ARPES), which provides access not only to the bands' onset energy, but to the whole *energy vs parallel momentum* dispersion of occupied GNRs states (as displayed in Fig. 4b) [34]. Being an ensemble averaging technique, domains of equally oriented GNRs are needed to get k-resolved ARPES data. This has been managed by growing the GNRs on vicinal surfaces that exhibit a periodic array of steps, which act as templates and drive the uniaxially oriented growth of GNRs along the terraces direction. Slight deviation between the value of the valence band maximum in flat and vicinal surfaces are found due to the different substrate´s work functions.

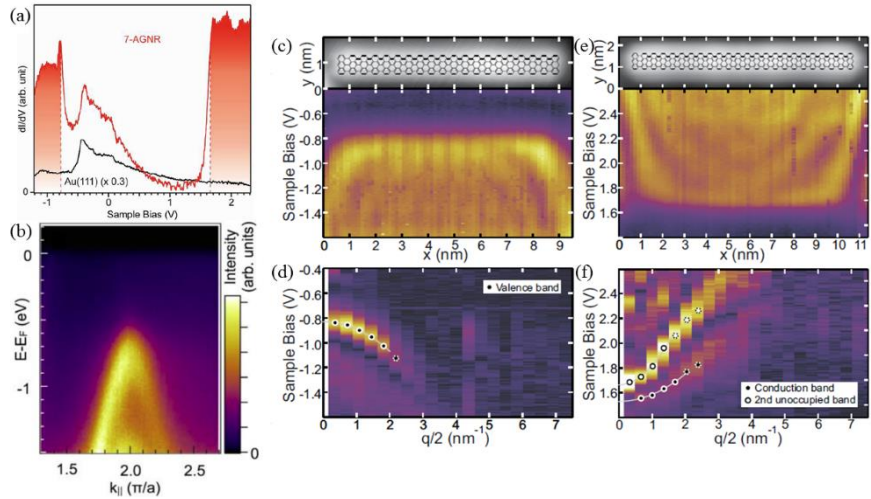

**Fig. 4** Electronic structure of 7-aGNRs. **a** dI/dV spectrum (red) of a 7-aGNR grown on Au(111). The onsets of valence and conduction bands are indicated by the shaded areas. Electronic features in the gap are due to the Au surface state (visible in the black spectrum). **b** ARPES intensity plot recorded for 1ML of 7-aGNRs grown on Au(788) recorded along the ribbon axis. The VB onset is found at -0.7 eV below the Fermi level. **c** STM image and plot consisting on equidistant dI/dV spectra of occupied states recorded along the edge of a 7-aGNR. **d** Line-by-line Fourier transform of **(c)** including a parabolic fit of the VB. The same is done for the CB on a longer ribbon in **(e-f)**. (a) Reprinted with permission from [51]. Copyright (2017) American Chemical Society. (b) Reprinted with permission from [34]. Copyright (2012) American Chemical Society. (c-f) Reprinted figure with permission from [50]. Copyright (2015) by the American Physical Society.

## *3.2 Chemical structure of Graphene nanoribbons*

*X-ray photoemission spectroscopy* (XPS) is the technique of choice to study the chemical composition of a material. In the case of GNRs, it has been used to shade light on the chemical mechanisms underlying the three basic steps in GNRs' formation by on-surface synthesis. After room temperature deposition of precursor molecules on the substrate, the shifts of the core levels (as Br3d and C1s) are measured as a function of increasing sample temperature in different experiments (see Fig. 5 for an example). This allows obtaining invaluable information about the whole reaction process including the step sequence, the halogen desorption from the surface and the associated threshold temperatures [25,30,58].

The chemical structure of GNRs has been imaged in UHV with STM and with atomic resolution by *non-contact atomic force microscopy* (nc-AFM). In molecular imaging, STM is sensitive to the density of states near the Fermi level, typically delocalized over the entire molecule. The correlation of bonding structure and STM contrast is thus not straightforward. However, simulations of STM images, reliably performed with DFT, are of great value for the interpretation of the STM data and finally allow identifying the structure and adsorption geometry of GNRs [18,43,46].

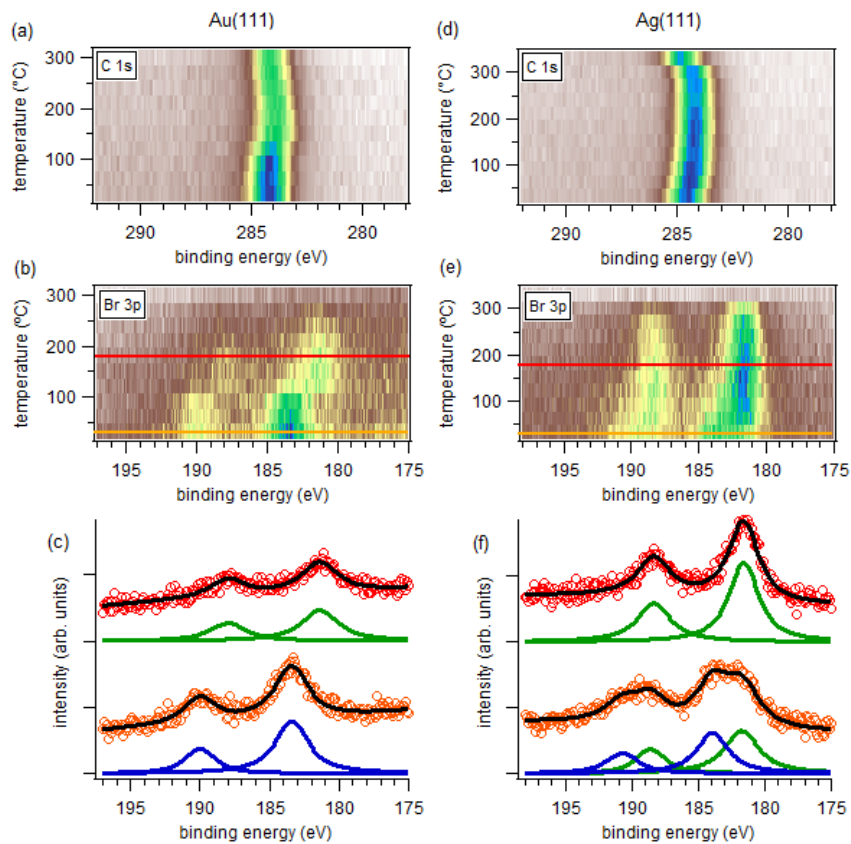

**Fig. 5** Photoemission spectra of the C 1s core levels of 2,2′-dibromo-9,9′-bianthracene deposited on (**a**) Au(111) and (**d**) Ag(111) held at room temperature and their evolution as a function of sample annealing temperature. Similar measurements of the Br 3p core levels are shown in panels (**b**) and (**e**). Panels (**c**) and (**f**) depict Br 3p spectra, together with their associated fits (blue and green lines correspond to organic and metal-bound Br components, respectively), of two representative temperatures marked with the colored lines in (**b**) and (**e**), respectively. The spectra are shifted along the intensity axis for better comparison. Reprinted with permission from [25].

The AFM technique allows, under specific conditions, sensing the short range forces between the probe and the measured sample, whereby the signal becomes sensitive to the total charge distribution on the adsorbates (which is largest on atomic sites and along chemical bonds). Although not strictly required [59,60], such imaging is most commonly performed at low temperature (~5 K), in frequency modulation and non-contact modes, by keeping an inert probe (typically functionalized with a single CO molecule) close enough to the sample [61]. Doing so and tracking the changes in the frequency shift of the oscillating cantilever force sensor in a plane above the GNR, it has been possible to precisely resolve the bonding structure of pristine and doped GNRs (Fig. 3b,c,d,k) [40,41,42,48], as well as defects [40] and atomically sharp junctions [29].

Besides its imaging capabilities, nc-AFM has been used also to proof fundamental properties of GNRs as their *structural superlubricity* on gold surfaces [62]. GNRs with lengths between 5 and 55 nm could be moved laterally by the force sensor tip with static friction forces between 2 and 200 pN. Such ultralow forces arising in the sliding motion of GNRs pave the way for creating frictionless coatings.

## *3.3 Detection of vibrational modes*

The GNRs vibrational structure predicted by theoretical studies, based on first principle methods, is characterized by graphene-like and intrinsic modes [63]. The graphene $E_{2g}$-like (or G) mode is induced by the relative motion of neighboring atoms and tends to the value of 1580 cm$^{-1}$ of the $E_{2g}$ in graphene as the GNRs' width increases. A localized mode is found at 3000 cm$^{-1}$ which is the typical vibration of the C-H bond. The most prominent peak in the low frequency range (200-800 cm$^{-1}$) is the *radial breathing mode* (RBLM) due to the outward motion of the nanoribbon's edge atoms by keeping the central atoms at rest (inset in Fig. 6b). This is the most representative vibrational mode of both armchair and zig-zag GNRs and since its frequency is roughly proportional to the inverse square root of the GNR's width, it can allow identifying the types of ribbons present on a surface.

Experimentally, the frequencies related to such vibrational modes of GNRs have been determined by Raman spectroscopy. Raman spectroscopy measures the inelastic scattering of photons by phonons, usually by means of a laser as photon source tuned in a spectral range between the near infrared to the near ultraviolet. Raman spectra measured on a layer of 7- and 9-aGNRs grown on Au(111) identified the different modes predicted by theory, as the RBLM mode at 396 cm$^{-1}$ for 7-aGNRs and 312cm$^{-1}$ for 9-aGNRs, but also the D peak (related to disorder) in graphene (Fig. 6 a,b) [18,45]. Raman peaks showed a strong dependence on the excitation source wavelength so that the RBLM of 9-aGNRs is depleted by incident green laser irradiation (2.33 eV) but enhanced by infrared light (1.58 eV), the opposite occurs for 7-aGNRs. According to tight binding calculations coherent radial-breathing-like phonons in GNRs are excited for photoexcitation near the optical adsorption edge [64]. In fact, in the case of 7-aGNRs, aligned uniaxially on Au(788), the size of the optical gap is 2.1 eV as determined by reflectance differ-ence spectroscopy measurements (RDS) [56]. Such value is then close to the green light excitation while infrared lies within the gap.

Raman is a powerful technique for the characterization of GRNs under several aspects: (i) it does not need ultra-high-vacuum conditions to be operated; (ii) it can be used to detect GNRs vibrational fingerprints also when they are deposited or grown on insulating surfaces; (iii) it allows identifying the orientation of the GNRs edges and the GNRs width [65-68]; (iv) it helps to evaluate the quality of a GNRs sample by quantifying the defect density via D band [65]. Ideally it will also allow to analyze GNRs length distribution.

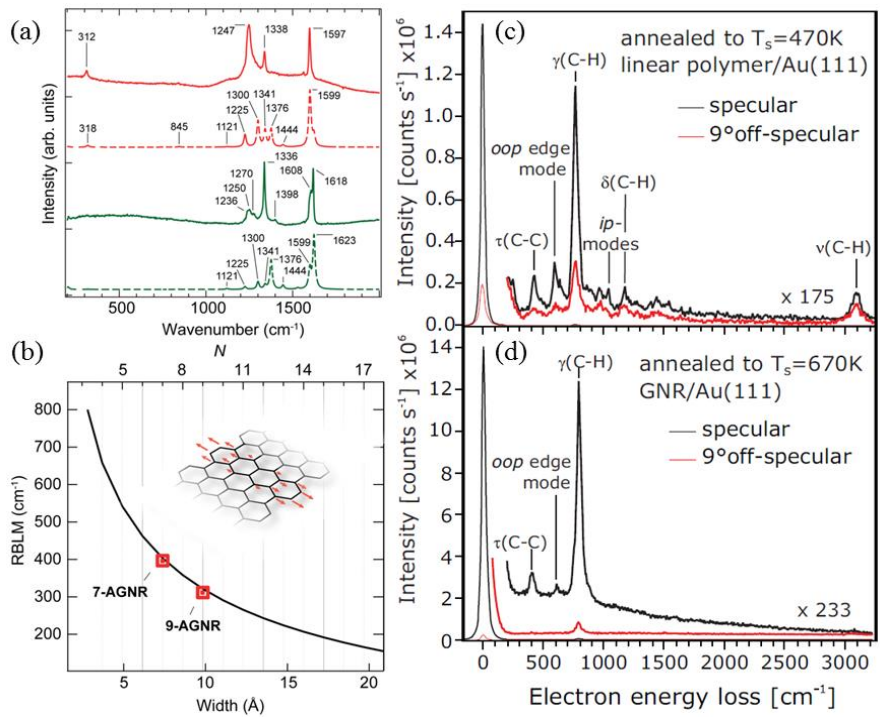

**Fig. 6 a-b** Raman characterization of 9-aGNRs grown on Au(111). **a** Experimental Raman spectra taken with two different excitation sources, 785 nm (solid red) and 532 nm (solid green) laser lines. Corresponding DFT based simulations done for photon energies of 1.58 eV (dashed red) and 2.33 eV (dashed green). **b** Theoretical dependence of the wavenumber of the RBML with the aGNRs width and experimental data for 7-aGNRs and 9-aGNRs. **c-d** Vibrational HREEL spectra corresponding to (**c**) linear polymer obtained after dehalogenation of DBBA at 470 K on Au(111) and (**d**) 7-aGNRs obtained after cyclodehydrogenation at 670 K. (a-b) Reprinted with permission from [45]. Copyright (2017) American Chemical Society. (c-d) Reprinted figure with permission from [52]. Copyright (2012) by the American Physical Society.

The "electron analogue" of Raman spectroscopy, naively, is high-resolution electron energy loss spectroscopy (HREELS). This technique uses the inelastic scattering of low energy electrons to measure the discrete vibrational energies of vibrational modes of adsorbates on a surface. Angle resolved HREELS has been used to follow and characterize the thermally activated steps of GNRs formation [52,55]. For GNRs well defined changes in the vibrational spectrum from the polymeric phase to the aromatic structure have been measured. The quenching of vibrational modes with dipole moments laying within the phenyl rings plane demonstrated the planarization of all the phenyl rings after cyclodehydrogenation (Fig. 6 c,d).

## 4 Tuning the electronic properties

Among the most attractive virtues of graphene nanoribbons is the remarkable dependence of their electronic properties on the detailed GNR structure. As a consequence, even minimal structural changes brought about in a controlled way through the synthesis process, allow tuning the nanoribbon´s electronic properties over a wide range. In the following we review some of the different routes that have been proposed and explored to control the atomic structure of graphene nanoribbons, as well as their effect on critical electronic properties as for example the band gap, band dispersion or energy level alignment.

### *4.1 Tuning through edge orientation*

As readily discussed in the introduction in the light of Clar´s theory, the edge orientation has a critical effect on the nanoribbon´s electronic properties. On a more quantitative basis, we will now discuss some of the details in reference to the electronic properties of 2D graphene. Tight-binding calculations have been successfully used to describe with reasonable accuracy the electronic properties of 2D and nanostructured graphene. Electrons involved in gaphene´s σ-bonds are more strongly bound and thus the main interest for optoelectronic processes arises from the delocalized π-electrons. Consequently, each atom can be simply characterized by a single $2p_z$ orbital. Applying the tight-binding Hamiltonian at the nearest neighbor level to the two-dimensional honeycomb structure of graphene with its two atoms per unit cell (each forming an equivalent hexagonal sublattice, Fig. 7), the following dispersion relation is obtained [69]:

$$E(k) = st\sqrt{3 + 2\cos\left(\frac{\sqrt{3}k_x a}{2} + \frac{k_y a}{2}\right) + 2\cos\left(\frac{\sqrt{3}k_x a}{2} - \frac{k_y a}{2}\right) + 2\cos(k_y a)} \qquad (1)$$

where $s = \pm 1$ (-1 for the valence and +1 for the conduction band), $t$ is the hopping integral and $a$ is graphene´s lattice constant. Near the $\Gamma$ point, both valence and conduction bands depend quadratically on $k_x$ and $k_y$. At the M points there is a saddle point in the energy dispersion. The most interesting point is around the K points, where the bands show a linear dispersion in so-called Dirac cones that display electron-hole symmetry (Fig. 7c) [69,70]. That is, this first order tight-binding calculation already predicts the semimetallic, gapless nature of graphene, as well as the linear dispersion around the K points in a massless electron behavior.

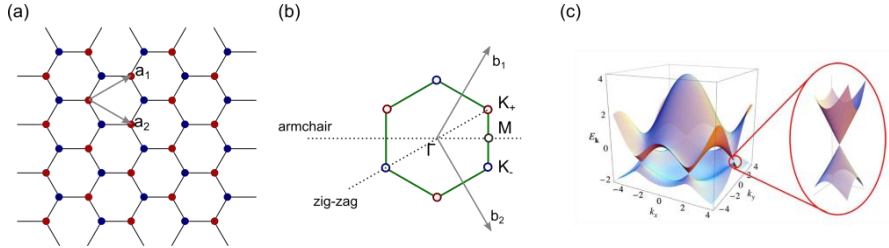

**Fig. 7 a** Graphene lattice and its primitive vectors $a_1$ and $a_2$. **b** Reciprocal space of Graphene and its reciprocal primitive vectors $b_1$ and $b_2$. The Dirac points are located at K±. Dashed lines indicate the growth directions of armchair and zig-zag GNRs in the reciprocal space. **c** Band structure of graphene. The close up shows the linear behavior of the bands close to the Dirac points. The slope of these cones correspond to the Fermi velocity $v_F$. (c) Reprinted figure with permission from [70]. Copyright (2009) by the American Physical Society.

Applying the same theoretical analysis to GNRs of arbitrary edge orientation, the following boundary condition is set: because H atoms along the edges do not contribute any π-electron, the GNR´s π-electron wave functions on them vanish. In aGNRs, this boundary condition is fulfilled by eigenstates of graphene with perpendicular momenta $k_\perp$ that allow standing waves perpendicular to the aGNR axis. A detailed derivation is provided in reference [69], resulting in the discrete values

$$k_\perp = \frac{r}{N+1}\frac{2\pi}{a}, \quad r = 1, 2, 3, ... N \quad (2)$$

The associated aGNR´s band structure is thus simply a combination of slices across graphene´s band structure at $N$ equidistant $k_\perp$ values. This is shown by way of example in Fig. 8 for 9-aGNRs. The frontier bands stem from the cut closest to graphene´s Dirac point at K. Depending on the nanoribbon´s width, the properties can thus vary from a metallic to a semiconducting behavior. The former occurs with aGNRs of the $N_a=3p+2$ family, which always present one cut right through the K point, while the former applies to the other two subfamilies. At this point it is important to note, however, that this tight-binding derivation is based on equiva-lent carbon atoms throughout the nanoribbons. Taking into account that carbon atoms at the edges are saturated with hydrogen, such equivalence is lifted and the resulting edge distortion actually results in the opening of a band gap also for the 3p+2 family [11,71], as obtained also from more refined *ab-initio* calculations [11,71] and as readily proved experimentally with 5-aGNRs [43]. Further experimental examples reporting the electronic properties of different atomically precise aGNRs and confirming the great variety of band gaps as a function of their width will be discussed in the next section.

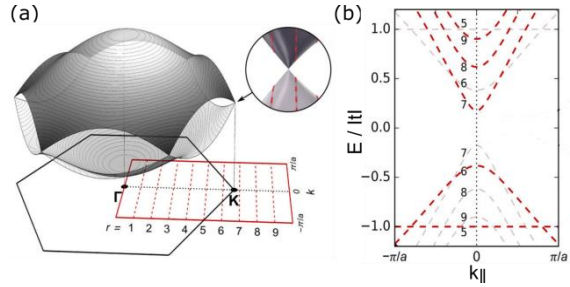

**Fig. 8 a** First Brillouin zone of graphene (black hexagon) and its π-band structure (black/gray surfaces) with inset displaying bands near the Dirac cone. Also shown are the first Brillouin zone (red rectangle) and cutting lines (red dashed lines) for infinite length 9-AGNRs. **b** Energy bands of empty and occupied states corresponding to cutting lines r = 5, 6, ..., 9 in (a). Reprinted with permission from [45]. Copyright (2017) American Chemical Society.

The case of zGNRs is more complicated. As derived in reference [69], the wave functions satisfying the boundary conditions have transverse momenta displaying a dependence on the parallel momentum such that

$$\sin(k_\perp N) + g_k \sin[k_\perp(N+1)] = 0, \qquad g_k = 2\cos\left(\frac{k_\parallel}{2}\right). \qquad (3)$$

That is, most of the bands can still be obtained from slices of graphene's band structure, but those slices are now curved, as displayed by way of example with the black solid lines in Fig. 9a for a zGNR with $N_z = 4$ (4-zGNR). It is immediately obvious that under this tight-binding approximation zGNRs are gapless because one of the slices always goes through graphene's K point. This is also evident from Fig. 9b, which displays the band structure associated with these ribbons. It is interesting to see in Fig. 9b that valence and conduction band join into a degenerate flat band above a critical $k_\parallel$ value $k_c$. It coincides with the reciprocal space region where only $N$-1 slices are present instead of $N$ (Fig. 9a), because no $k_\perp$ are found as solutions to equation (3). In that region, the states giving rise to the flat bands can be understood as imaginary solutions to equation (3) at $k_\perp = 0$ and $k_\perp = \pi/a$, as displayed with dotted red lines in Fig. 9a. The wave functions associated to those solutions are localized at the nanoribbon edges and strongly decay toward the interior of the zGNRs [69,72,73]. It is best observed that these edge states can-not be derived from slices in graphene´s band structure comparing its projection along the ΓK direction (Fig. 9d) with the band structure of wide z-GNRs (Fig. 9c), which has a large density of slices and could thus be expected to be comparable to that of graphene. Such comparison immediately makes obvious that the flat bands associated to the edge states are not present in 2D-graphene, but that they are a property of zGNRs intrinsic to their zigzag edges.

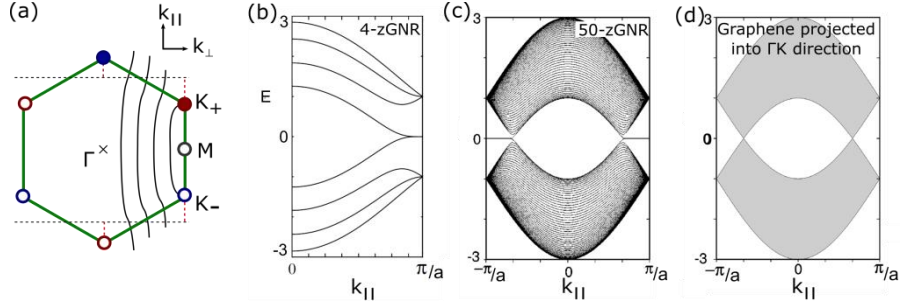

**Fig. 9 a** Reciprocal lattice of graphene showing the discrete values of $k_\perp$ for N = 4. The cuts are not linear, in contrast with aGNR. Also note the change in orientation of $k_\perp$ and $k_\parallel$ with respect to Figure 8. The red dotted lines mark the regions where the additional edge states appear that cannot be traced back to the band structure of graphene. The reciprocal unit cell of a 4-zGNRs is also shown in black dashed lines. **b** Band structure of a 4-zGNR, revealing the degenerate edge states as $k_\parallel$ approaches π/a. **c** Band structure of a 50-zGNR revealing the flat bands of the edge states even more clearly. **d** Graphene´s band structure projected into GK direction, showing the absence of the flat bands of zGNRs. π/a indicates the boundary of the first 1D Brillouin zone of the GNR (black dotted lines in (a)), where the zone-folding technique should be applied. (b) Reprinted figure with permission from [73]. Copyright (1996) by the American Physical Society. (c) and (d) Reprinted figure with permission from [72]. Copyright (1999) by the American Physical Society.

The strong density of states at the Fermi level caused by the edge states should induce either a lattice distortion via electron-phonon interactions or magnetic polarization via electron-electron interactions [69]. In the case of GNRs the latter is favored and drives the spin polarization of the edge states. Edge sites along a zGNR are made up by atoms of the same sublattice on one side and of the other sublattice on the opposite side. Thus, the edge state being nonzero only on one of the two sublattices at each edge, the magnetic moment selectively increases on that sublattice with a ferromagnetic spin configuration and decays toward the ribbon interior. The opposite edge sites belong to the other sublattice and display magnetization with opposite spin. The edge states are thus not only spin polarized ferromagnetically along each edge, but additionally antiferromagnetically coupled across the ribbon [69]. Notably, the electron-electron interactions driving the magnetization concurrently open a band gap $\Delta^0$ between the edge states, deterring zGNRs from truely being gapless structures. This can be observed in the calculated band structure of zGNRs with and without electron-electron interactions in Fig. 10b. While tight-binding density of states has only one van Hove singularity relat-ed to the edge state´s flat bands at E = 0, including the electron-electron interaction U term in a mean field Hubbard model results in two pairs of density of states peaks split by $\Delta^0$ and $\Delta^1$ (Fig. 10b). In particular, $\Delta^0$ relates to the antiferromagnetic correlation between the two edges, while the larger splitting $\Delta^1$ relates to the ferromagnetic correlation between the spins of the most strongly localized states at $k_\parallel = \pi/a$ in the same edge. It should be mentioned here that in the nearest-neighbor approximation the electron-hole symmetry remains unchanged, but it is lifted as a next-nearest-neighbor hopping term is included in the Hubbard model [74]. The opening of a bandgap has been confirmed on atomically precise 6-zGNRs synthesized on Au(111) that were subsequently transferred onto NaCl bilayer islands, indeed revealing under such circumstances relatively large splitting values of $\Delta^0$ = 1.5 eV and $\Delta^1$ = 1.9 eV [42]. Although with somewhat disparate values, similar conclusions have been drawn also from experiments on zGNRs with worse defined and less perfect edges obtained by differen methods like hydrogen etching of 2D-graphene [13] or scanning probe lithography [15].

From a structural point of view, chiral ribbons combine zigzag and armchair segments. From an electronic point of view the zigzag segments are most determining, providing chiral ribbons with low band gaps and spin-polarized edge states. However, these properties vary with the chiral angle ($\theta$, Fig. 2) in a relatively smooth way in between the two limits of zigzag ($\theta=0°$) and armchair graphene nanoribbons ($\theta = 30°$). In the infinite-width limit, the band structures of chiral GNRs can be obtained from a continuous rotation of the band structure of graphene and the average density of edge states per edge length $\rho_0$ at the Fermi level is given by [75]

$$\rho_0(\theta) = \frac{2}{3a} cos\left(\theta + \frac{\pi}{3}\right). \qquad (4)$$

As discussed above, the large $\rho_0$ is responsible for the magnetization of GNRs. And as evidenced in equation (4), $\rho_0$ is highest for lowest chiral angles and vanishes as the armchair angle is approached [74,76]. This has the important implication that the magnetic moment per length unit increases as the chiral angle is reduced. Tight-binding calculations of chiral GNRs of finite width as a function of the chiral angle confirm this picture [76]. For GNRs with edge orientations close to aGNRs the edge states associated to the flat bands at the Fermi level are almost completely suppressed (Fig. 10a). Instead, in ribbons with edge orientation close to a zGNR the flat edge state band extends over the whole 1D Brillouin zone and becomes multiple degenerate due to band folding (Fig. 10a) [76]. As in zGNRs, electron-electron interactions split the degeneracy of the flat bands also in chiral GNRs (Fig. 10c,d). As discussed earlier, each pair of peaks arising in the density of states has a different nature, whereby $\Delta^0$ and $\Delta^1$ relate to the magnetic coupling across the nanoribbon and along its edge, respectively. As can thus be intuitively expected, for a given GNR width $\Delta^0$ hardly changes with the chiral angle. Instead, $\Delta^1$ decreases in a similar way to the net magnetic moment on each edge, as the chiral angle increases and approaches the armchair orientation [76].

Lastly, it should be remarked that the density of edge states as a function of chiral angle depends at the same time on the nanoribbon width. Beyond the visualization through the Clar sextet theory outlined in the introduction, this effect has been computed calculating the number of states at E ≈ 0, normalized to the total number of states in the GNR [73,77]. Calculations reveal that there is a critical width below which there is no notable edge state density. Importantly, that critical width decreases with decreasing chiral angle (as the edge orientation approaches that of zGNRs) [73,77]. In other words, for a given width there is a minimum number of zigzag sites (in relation to armchair sites) along the ribbon´s edges to host a notable edge state density. The required ratio of zigzag to armchair sites becomes higher for narrower GNRs.

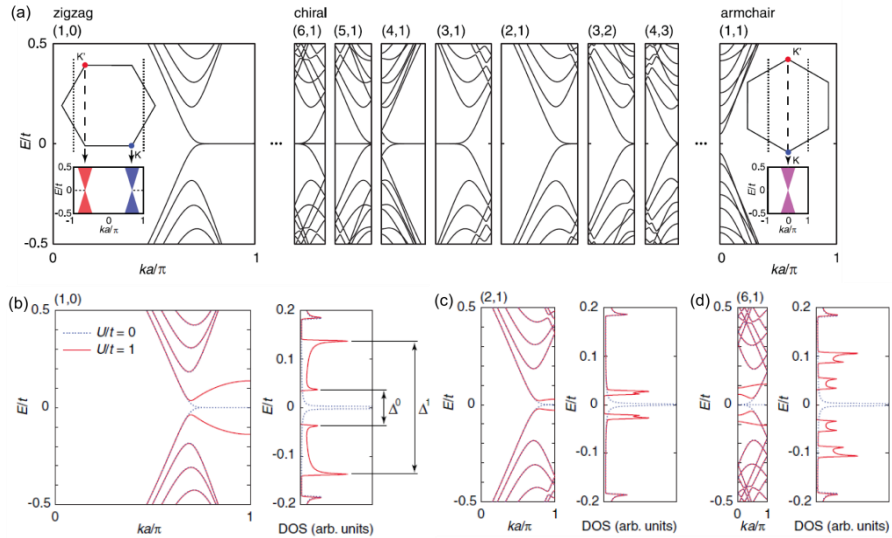

**Fig. 10 a** Band dispersion obtained from tight binding calculations of GNRs of different orientation, from zigzag, through chiral, to armchair. The closer to the armchair direction, the lower the density of states of the flat bands at E=0, until a band gap opens for armchair ribbons. Similar calculations for zigzag and chiral ribbons are shown in **b-d** upon addition of electron-electron interactions, revealing the opening of a bandgap also for these orientations. Reprinted figure with permission from [76]. Copyright (2011) by the American Physical Society.

It is important to observe at this point that the predictions described above for the electronic structure of chiral GNRs still lack experimental proof on atomically precise chiral GNRs. To date, only the precise synthesis of narrow (3,1)-GNRs has been published [25,31,32], and only few details of its electronic properties are yet known [65]. There are few other works reporting the electronic structure of chiral graphene nanoribbons with worse defined edges obtained from unzipped carbon nanotubes [12] or chemical vapor deposition [14]. In either case the opening of band gaps through the magnetism-driven splitting of the edge states have been confirmed by scanning tunneling spectroscopy measurements.

## *4.2 Tuning through width control*

Besides the edge orientation, the electronic structure of graphene nanoribbons also displays a marked dependence on their width. It is most straightforward to understand this dependence for armchair graphene nanoribbons, which is furthermore the family with the largest number (albeit still very limited) of successfully synthesized examples on which to experimentally corroborate such behavior. We remind that the band structure of *N*-aGNRs can be pictured as a combination of 1D-cuts across graphene´s band structure at *N* equidistant $k_\perp$ values, namely at $k_\perp = \frac{r}{N+1}\frac{2\pi}{a}$, *r* being an integer between 1 and N. Keeping in mind that graphene´s vanishing band gap and the massless electron behavior occur right at the K-point, the closer those 1D-cuts get to the K-point, the lower the aGNR´s band gap and the effective masses of its frontier bands will be. That is, some of the aGNRs´ attributes can be readily understood from simple geometric considerations: along the 1D-projection, K and K´ are located at $k_\perp = \frac{1}{3}\frac{2\pi}{a}$ and $k_\perp = \frac{2}{3}\frac{2\pi}{a}$ (Fig. 8a). It is the 3 in the denominator that accounts for the presence of three aGNR families, depending on whether their width in terms of dimer lines *N* is equal to 3p, 3p+1 or 3p+2 (p being an integer). There is always a cut going directly through the K-point for the 3p+2 family. This family thus has the lowest band gap and effective mass, these values being higher on either of the other two families for which no cuts across the K-point are present. However, the wider the GNR, the smaller the spacing between the cuts and the closer they get to the K-point. This determines the inverse proportionality of band gap and effective mass with nanoribbon width within each of the families, although if the width increases atom by atom it would imply changes between families that could result in an increased band gap and effective mass as well.

It is worth remarking here that, under the assumption of equivalent carbon atoms throughout the GNR, tight binding calculations predict the 3p and 3p+1 families to follow a very similar trend. However, this assumption is inaccurate the closer the atoms are to the edge, most evidently for the hydrogen saturated carbon atoms. Taking into consideration bond length distortions and the associated changes in the hopping constant, important changes appear in the resulting band gap calculations, as is a clear splitting of the 3p and 3p+1 bandgap trends, as well as the opening of a bandgap for the 3p+2 family. These findings coincide with more refined *ab-initio* calculations revealing the band gaps of the three different families to order according to $E_g$ (3p+1) > $E_g$ (3p) > $E_g$ (3p+2) [11,71].

Besides tight-binding [69,11,71], several other methods have been used to quantitatively describe the width-dependent band gap in aGNRs, like for example extended Hückel theory (EHT) [78], density functional theory (DFT) [11,79], or the GW approximation [79]. At this point, it is important to remind that, to date, all the experimentally reported band gaps of atomically precise GNRs are from surface-supported ribbons. While less important on weakly interacting substrates, if more reactive surfaces are used or reactive functional groups are added to the GNR structure, strong GNR/substrate hybridizations may occur. Under such circumstances the band gap will be substantially affected and the GNR´s electronic properties may bear little resemblance with those of free-standing GNRs. In addition, the band gaps of GNRs have been most commonly characterized by ionizing techniques such as STS [34,43,45,46] or a photoemission/inverse photoemission combination [35]. This implies on the one hand that it is not ground states but quasiparticle energies which are probed, and on the other hand that the surrounding dielectric medium (both the GNR itself as well as the substrate) can have a substantial effect on those state´s energies through polarization-induced screening. Among the various theoretical methods mentioned above, the intrinsically present electron-electron interactions are best taken into account in the GW approximation. It is, however, computationally very intensive and has thus only been applied for free-standing GNRs. As a result, calculated GW band gaps still overestimate the experimental ones [34]. Best results have been obtained when combining the GW approximation with a semiclassical image-charge model to account for substrate screening [80]. The screening-derived band gap renormalization varies with the substrate and the GNR, being in the 1 eV range for aGNRs on Au (Fig. 11) and slightly lower for aGNRs on a NaCl bilayer on Au [80]. Its combination with the calculated GW band gaps ultimately provides a remarkably good agreement with currently available experimental values for 5 [43], 7 [50], 9 [45] and 13-aGNRS [46] (Fig. 11).

The effective mass is another important figure whose relevance stems from its inverse relation to the charge carrier mobility [81], a critical parameter for potential applications. However, only few studies have characterized effective masses in GNRs. From the experimental side, there are measurements for 7-aGNRs and 9-aGNRs. For the valence band of the former, very disparate values have been reported, ranging from ~0.21 $m_0$ (measured by ARPES [34,82]) to 0.41 ± 0.08 $m_0$ (measured by STS [50]) and even 1.37 $m_0$ (measured by two photon photoemission [53]). For the valence band of the latter, better matching values of 0.09 ± 0.02 $m_0$ and 0.12 ± 0.03 $m_0$ have been reported from ARPES and STS measurements, respectively [45]. Interestingly, there have been different works proposing analytical relations between the effective mass and other GNR parameters. For example, Arora et al. proposed the effective mass to be linearly dependent on the GNR band gap in a similar way as that claimed for carbon nanotubes: $m^* = (E_g / 11.37)\, m_e$, with $E_g$ given in eV [83]. In turn, Raza et al. related it to the aGNR width W through $m^* = (0.091 / W)\, m_e$ for the 3p family, $m^* = (0.160 / W)\, m_e$ for the 3p+1 family and $m^* = (0.005 / W)\, m_e$ for the 3p-1 family (W being the distance, in nanometers, between the C atoms on either edge of the ribbon, based on a C-C bond length of 1.44 Å) [78]. In the following we provide Table 1 comparing the exper-imentally reported effective masses to the values predicted by either of the models mentioned above. It can be seen that there is an excellent agreement between predictions and experiments, best of all when taking the experimental values obtained from ARPES measurements.

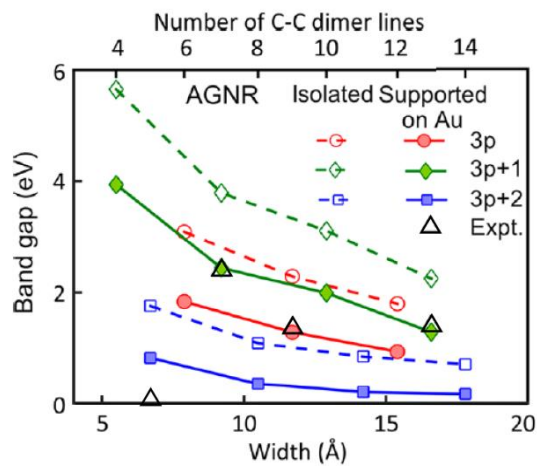

**Fig. 11** Calculated bandgap energies for isolated and Au-supported aGNRs of different width for each of the three sub-families (3p, 3p+1, 3p+2). Calculations are with the GW method, further adding substrate screening for the supported ribbons. Superimposed we find the currently available experimental values of Au-supported aGNRs obtained by scanning tunneling spectroscopy, providing good agreement. Adapted with permission from [80]. Copyright (2016) American Chemical Society.

**Table 1** Summary of the calculated and experimentally obtained values for the effective mass of AGNR´s valence bands.

|  | Arora et al.[83] | | Raza et al. [78] | Experiment | | |
|---|---|---|---|---|---|---|
| width (N) | From calc. $E_g$ | From exp. $E_g$ | | ARPES | STS | 2PPE |
| 7 | 0.154 | 0.21 | 0.21 | 0.21 [34] 0.23 [82] | 0.41± 0.08 [50] | 1.37 [53] |
| 9 | 0.074 | 0.12 | 0.09 | 0.09±0.02 [45] | 0.12±0.03 [45] | --- |

Experimental $E_g$ values for the m* calculation with Arora´s relation have been taken from refs. [50] and [45] for 7-AGNR and 9-AGNR, respectively.

The width dependence of the electronic properties of GNRs does not only affect armchair oriented ribbons, but also ribbons with other orientations. As described in the previous section, the edge state magnetization in zigzag and chiral GNRs drives the opening of a band gap $\Delta^0$ and an additionally split resonance $\Delta^1$ related to the inter-edge and intra-edge magnetic coupling, respectively (Fig. 10) [69,74,76]. Thus, as expected from their respective nature, $\Delta^1$ hardly varies with the ribbon´s width. Instead, $\Delta^0$ decreases with growing width, although in the zGNR case in a monotonic way and faster than for aGNRs [74-76]. This qualitative behavior of $\Delta^0$ and $\Delta^1$ is at least what results from calculations on free standing zGNRs. However, while the same trend holds for $\Delta^0$ also in calculations of gold-supported ribbons, in that case $\Delta^1$ shows a surprising increase with ribbon width [80]. Experimental confirmation on atomically precise zGNRs of different width is still missing, since only 6-zGNRs have been synthesized to date with atomic precision [42]. Nevertheless, closely related measurements have been performed on the edge states localized at the zigzag ends of 7-aGNRs. For increasing 7-aGNR length, which increases the distance between their ends (in analogy to a width increase in zGNRs), the gap between edge states hardly changes. Because the minimum separation measured in that study is ~ 3 nm, this observation is in agreement with a rapidly converging band gap at even smaller length scales, further supported with calculations [84].There are also closely related measurement on zGNRs obtained from scanning probe lithography and hydrogen etching. The latter shows band gaps monotonously decreasing with increasing zGNR width until saturating at widths around 4 nm, in reasonable agreement with the predictions described above [13]. The former shows a smoothly decreasing band gap with increasing width until a threshold value (~7 nm) above which the gap suddenly drops to zero, proposed to be related to a sharp antiferromagnetic semiconductor to ferromagnetic metal transition [15]. Nevertheless, both studies are based on zGNRs with poorly controlled and defined structures as compared to the atomically precise nanoribbons that are the focus of this chapter, and thus possibly affected by the additional disorder.

Beyond the gap, also the edge state magnetization, which is one of the most coveted electronic properties of zGNRs, varies with GNR width. In particular, it increases as the width grows, rapidly reaching its saturation value at widths around $N_z = 10$ [69,74]. However, it is important to consider also the relative weight of those magnetic edge states on the total density of states of the nanoribbon. Initially they increase with increasing $N_z$, but after a peak around $N_z = 7$ it ends up decreasing at a rate approximately proportional to $1/N_z$ (Fig. 12a) [73]. This effect shows how the significance of the special edge state disappears as the ribbon grows infinitely wide and becomes 2D graphene (Fig. 12b-d).

Chiral ribbons also display a dependence of their electronic properties with their width. By way of example, and as readily discussed briefly in the previous section, the edge state density in chiral GNRs is absent for widths under a critical value [73,77]. That critical width depends on the chiral angle and decreases as the edge orientation gets closer to the zigzag direction. Regarding the band gap, it does not decrease monotonically with increasing width as in zGNRs, but displays a periodic modulation on top of its overall reduction, reminiscent of aGNRs [77,85]. However, it is worth mentioning that this overall band gap reduction is faster in chiral ribbons than in aGNRs. Describing the band gap dependence as $E_g \propto 1/N^\beta$, $\beta$ increases from 1 for aGNRs, to larger values as the chiral angle is re-duced (e.g. β=1.431 for θ=23.41°; β=2.908 for θ=16.1°; or β=3.758 for θ=12.73°) [77]. In contrast, the overimposed band gap modulation becomes less pronounced for lower chiral angles [77,85].

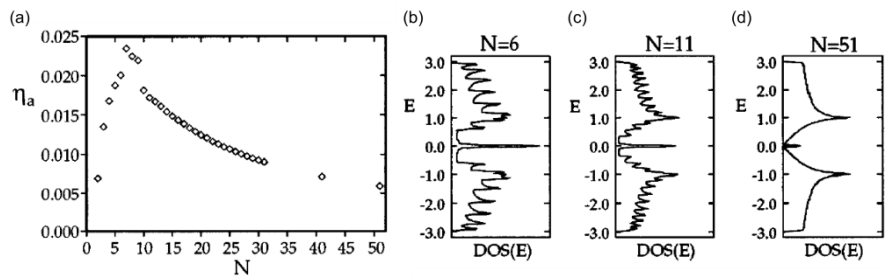

**Fig. 12 a** Flatness index $\eta_a$, defined as the ratio between the number of states with E ≈ 0 and the total number of states in the ribbon, calculated as a function of zGNR width $N_z$. Density of states calculated for zGNR widths of $N_z = 6$ (**b**), $N_z = 11$ (**c**) and $N_z = 51$ (**d**), evidencing the decreasing significance of the edge states in the overall density of states for increasingly wide ribbons. Reprinted figure with permission from [73]. Copyright (1996) by the American Physical Society.

Lastly we would like to remind that the discussion above has been made considering quasi-infinite ribbons on which only the width is changed. However, atomically precise GNRs obtained by on-surface synthesis are finite, typically with lengths in the range from few nanometers to some tens of nanometers [25,35]. As for other conjugated molecules or polymers, their band gap scales inversely proportional to their length, saturating towards the "infinite" limit [9,84]. Thus, although for most applications the growth of GNRs with maximized length is desirable, the growth of monodisperse GNRs of limited length could be seen as an alternative way to tune and control their corresponding band gaps.

## *4.3 Tuning through doping*

To understand which effect might doping bring to semiconducting GNRs one can look to the effect of impurities on semiconductors [86,87]. Three scenarios might be naively distinguished in: (i) the formation of impurity levels in the semiconductor's bandgap; (ii) the appearance of an impurity band; (iii) a rigid shift of the whole band structure or modification of the bandgap. In the first case, at *light* impurities concentrations (≈ $10^{16}$ atoms cm$^{-3}$), n-type (or donor) impurities give rise to energy levels below the semiconductor conduction band while p-type impurities yield electron acceptor levels above the valence band. Doping influences the position of the Fermi level, that for n-type moves linearly towards the minimum of the conduction band with exponential increase of the doping concentration. In the second case at *moderate* impurity concentrations (≈ $10^{18}$ atoms cm$^{-3}$ for Si), neighboring impurities are so close that their wavefunctions overlap sufficiently to give rise to an impurity band. In *heavily* doped semiconductors ($\approx 2\times 10^{22}$ cm$^{-3}$ i.e. 3% for Si) the ionization energy of the impurity atoms falls to zero, i.e. the impurity band merges with the conduction (for donors) or valence band (for acceptors) and only one band is formed. In other cases of heavy doping, as in disordered alloys, the band gap can be changed, for example linearly narrowed by adding Al atoms in Ga$_{1-x}$Al$_x$As.

For carbon-based materials two doping schemes have been used: (i) *electrical doping* which does not alter the lattice structure or chemical composition of the material, as by adsorption of a gas or a metal; (ii) *chemical doping* which can be achieved by substitution of carbon (C) atoms with heteroatoms like nitrogen (N), boron (B), silicon or sulfur (S). The first type of doping has been reported for Li-doped 7-aGNRs [82]. Heavy doping of 7-aGNRs with a monolayer of Li lowers the valence band by 1.33 eV and brings the conduction band onset below the Fermi level, thereby causing a semiconductor-to-metal transition of the ribbons. Highest doping levels (reached upon saturation at around two monolayers of Li) reveal a transfer of up to 0.05 electrons per carbon atom. Importantly, such doping is accompanied by a quasiparticle bandgap renormalization from 2.4 eV to 2.07 eV, as well as by a dramatic increase of the conduction band´s effective mass.

The second type of doping has been successfully demonstrated on atomically precise GNRs by means of chemically substituted molecular precursors. The inclusion of N, S or B atoms in the GNR matrix corresponds to a heavy doping level of several at% but the resulting electronic structure of GNRs cannot be directly related to dopant concentration [88]. In fact, extra-electrons may remain in molecular orbitals localized around the dopant heteroatom instead of extending into the GNR π orbitals [10]. The appropriate design of reactants has allowed to precisely introduce dopant heteroatoms either on GNRs edges [36,55,89,90] or within their backbone [48,91]. Ultimately it is the precise combination of heteroatomic species and their localization which defines their effect on electronic properties like band gap [10] or energy level alignment.

In the first case two different situations have been realized. The substitution of C-H groups with N atoms in the edges of chevron type GNRs leads only to a rigid downshift of the conduction and valence band energies by 0.1-0.13 eV per N atom per precursor molecule unit, thus making it a stronger electron acceptor (n-doping of the ribbon) [55,89,90]. The decrease of the position of the valence band by increasing the number of N dopants in the precursor molecule has been also characterized with ARPES, STS and DFT for 3-aGNRs grown on stepped Au(788) as shown in Fig. 13 a-c [36]. The electron lone pair of those nitrogen atoms being not in conjugation with the GNR π-system, it does not further affect the density of states of the frontier bands, leaving critical parameters like the band gap or the band´s effective mass unchanged.

On the contrary the substitution of (C-H)$_2$ groups by S atoms in 13-aGNRs' edges allows hybridization of one of the heteroatom´s lone pairs with the delocalized π-bands. Thus, as predicted by ab initio simulation for free standing ribbons, the energy gap decreases 140 meV with respect to undoped ribbons [49]. Howev-er, dramatic changes on the energy level alignment are not induced by S atoms because their electronegativity difference with respect to C is smaller than in the case of N atoms.

Finally, B impurities have been successfully introduced in the 7-aGNRs' backbone, allowing a full conjugation of empty B $p_z$ orbitals with the ribbon's π-system [48,91]. According to calculations, B-doping induces the formation of a new acceptor band lying only 0.8 eV above the valence band. As a result, the band gap is strongly decreased with respect to that of pristine 7-aGNR (displayed in Fig. 13 d,e). Experimentally, the acceptor band and its distribution along the B-doped 7-aGNR backbone has been measured by STS, showing good agreement with calculations (Fig. 13 f,g). However, the position of the valence band (and consequently the experimentally determined band gap) has not been reported yet.

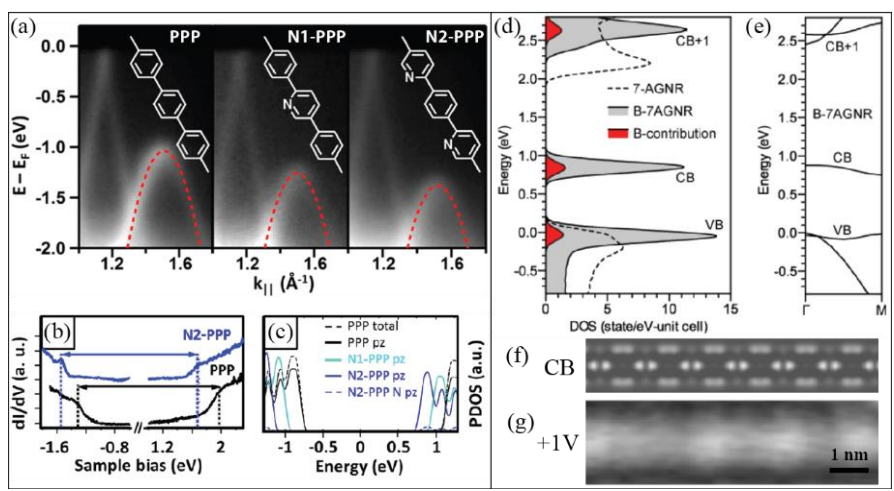

**Fig. 13 a** ARPES spectra of pristine and nitrogen doped poly-para-phenylene (PPP) polymers taken at $k_∥$ close to the valence band maximum. Red lines correspond to parabolic fits of the GNR´s bands revealing an unchanged effective mass of 0.19 $m_0$. **b** Differential conductance spectra performed on PPP (black) and N2-PPP (blue), displaying the rigid shift of valence and conduction bands with N-doping. **c** Computed projected densities of states (PDOS) of the different polymers, confirming again the rigid shift of valence and conduction bands upon N-doping and underlining the low contribution of the N atoms to the $p_z$ states forming the frontier bands. **d** Calculated total density of states (DOS) for pristine 7-aGNRs (dotted line), for B-doped 7-aGNRs (gray) and contribution from B-atoms to the DOS (red) using the GW approximation and including Au(111) substrate screening. **e** Calculated quasiparticle band structure of B-7aGNRs. **f** Calculated local density of states (LDOS) map of states at the conduction band edge 4 Å above the borylated ribbon. **g** Differential conductance map of B-7aGNRs taken at 1V. (a-c) Adapted with permission from [36]. (d-g) Adapted with permission from [91]. Copyright (2015) American Chemical Society.

## *4.4 Tuning through strain*

The dependence of the electronic structure of GNRs on the atomic structure of the ribbon orientation, the ribbon width and doping have been experimentally demonstrated by on surface synthesized GNRs. Strain engineering in semiconducting two dimensional materials offers the possibility to control their optical and electronic properties [92]. Thus another route to tune the band-gap of GNRs is to exploit strain. To understand the effect of strain on GNRs, it is necessary first to revise what occurs on strained graphene. Two types of strain are usually considered: uniaxial and shear (Fig. 14a,b). When graphene is stretched in one direction, it will shrink in the perpendicular one [93]. Tight binding and first principle calculations demonstrated that the effect of a uniform strain applied to the atomic lattice of graphene is to drive in the reciprocal space the position of the Dirac cone crossing ($E_D$) away from the K (K') points while maintaining the cone-like energy dispersion. This $E_D$ will follow a path that is different depending on the type of strain applied (shear or uniaxial along armachair or zigzag directions).

For aGNR shear strain modifies only slightly the band structure but uniaxial strain has a relevant effect. The one dimensional Brillouin zone of aGNRs is defined by electronic states with allowed $k$ values that lie on parallel lines ($k=r\pi/N+1$ with $r=1, 2, …,N+1$) (Fig. 14c). The three families of aGNRs characterized by their different width (*N=3p, 3p+1, 3p+2; p* being an integer) have different conditions for the crossing of their allowed k-lines with the K (K') points of the Brillouin zone (Fig. 14d). When uniaxial strain is applied $E_D$ moves away from K (K') along the strained Brillouin zone in a direction perpendicular to the k-lines (Fig. 14c). In this way when $E_D$ arrives in the middle of two lines, the energy gap will be maximum; when $E_D$ coincides with a k-line, then the gap will close. This finding explains why the energy gap of aGNR is modified in a periodic way with a zigzag pattern which is different for each family (Fig. 14b). Moreover, within each family, the wider the ribbon is, the closer the k-lines are. Thus, the maximum energy gap value will diminish.

For the case of zigzag GNRs, strain can slightly modify the band structure only if spin-polarization is taken into account. In this case uniaxial strain will increase the bandgap but shear strain will not have much influence.

To conclude, as observed in the case of graphene, GNR growth on substrates which are not lattice-matched might induce strain that eventually is released by buckling, fringes, ripples or blisters formation. Interface strain induced on GNRs synthesized on a surface has not been observed experimentally yet, but it might occur by increasing the interaction of the GNR with its support.

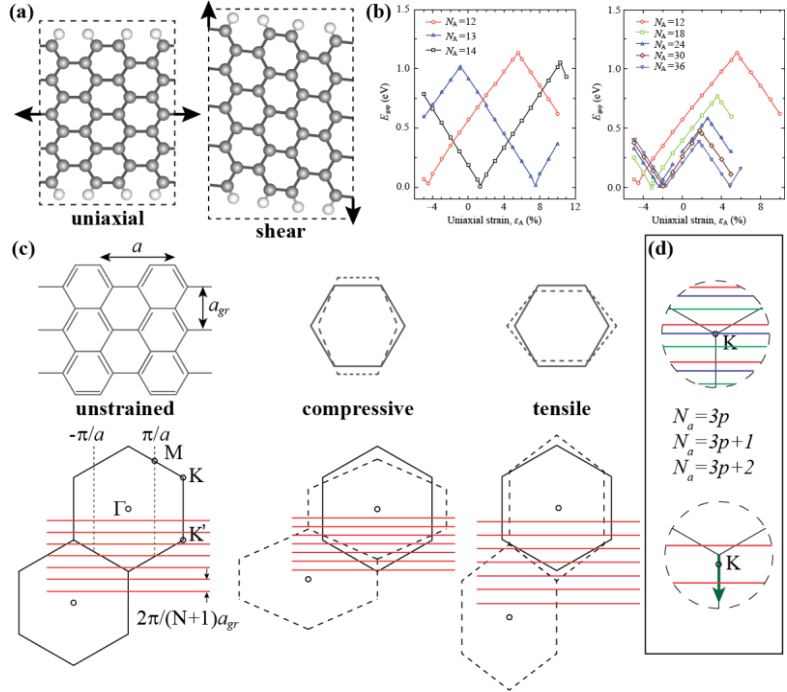

**Fig. 14 a** Schematic view of an aGNR under uniaxial and shear strain. **b** Calculated energy gap of aGNRs as a function of an uniaxial strain $\varepsilon_A$ on the left, for three ribbons belonging to the three families $N=3p$, $3p+1$ and $3p+2$ in the left and on the right, for different ribbons of the $N=3p$ type. **c** Schematic illustration of the effects of strain that induce (i) the deformation of the graphene Brillouin zone and motion of the K and K' points, (ii) a change of the distance between k-lines for allowed aGNRs electronic states and (iii) a different crossing of k-lines with the K and K' points. **d** Behavior of the allowed k-lines near the K point for the three aGNRs families. The green arrow indicates the shift of the Fermi point under uniaxial strain. Adapted with permission from [93].

## *4.5 Tuning through heterostructure formation*

Heterostructure formation has been successfully used with inorganic semiconductors to modulate their electronic properties, whereby sophisticated devices have been realized [94,95]. In analogy, the same concept has been proposed to be feasible with GNRs [96,97], for which one may combine segments with different properties like those described above (edge orientation, width, doping, strain), or also other parameters like edge terminations or edge structures. Experimentally, some examples have been readily demonstrated, creating atomically sharp and precise junctions by combination of different reactants (Fig. 15) [89,98]. An alternative way to create randomly distributed GNR heterostructures is by the use of only one type of reactant that can, however, react in two different ways [99].

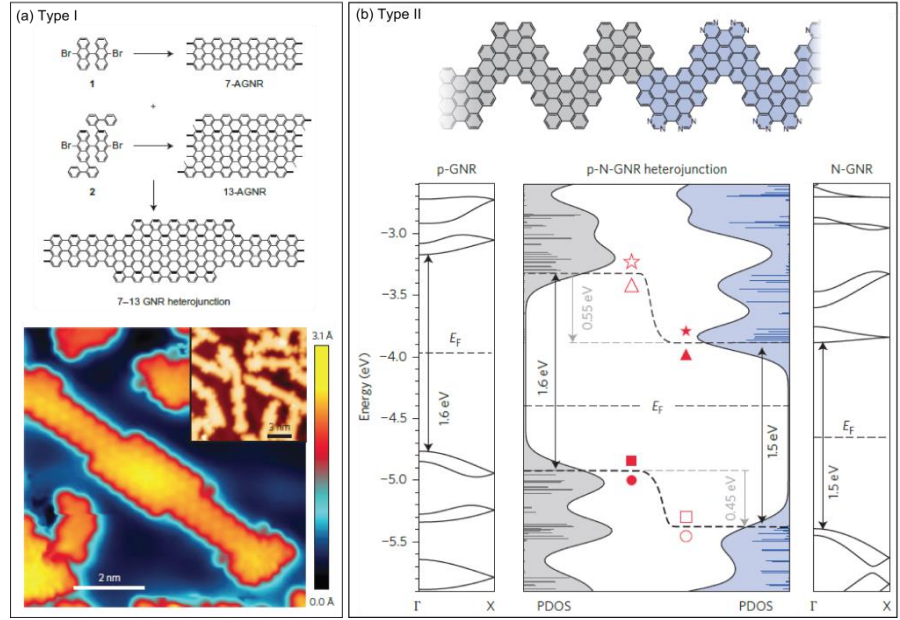

**Fig. 15 a** Schematic representation of the precursors leading to differently wide GNRs and an associated heterostructure, whose staggered bandgap can be classified as a type I heterojunction. Below, constant current STM images display the experimental realization of those heterostructures. **b** Schematic representation of the heterostructure resulting from a combination of pure hydrocarbon and N-doped precursors. Below the calculated electronic properties are displayed: band structure of pure p-GNRs (undoped precursors), of n-GNR (N-doped precursors), and of the density of states of the heterojunction displaying the roughly unchanged bandgap across the interface and the type II heterojunction energy alignment. (a) Reprinted with permission from [98]. (b) Reprinted by permission from Macmillan Publishers Ltd: Nature Nanotechnology [89], copyright (2014).

First to be realized was a type II heterojunction in which pure hydrocarbon reactants were combined with doped reactants that included nitrogen heteroatoms (Fig. 15b) [89]. As described in Sect. 4.3 and graphically displayed in Fig. 15b, the substitution along the GNR edges of C-H by N has little impact on the electronic band gap. In turn, nitrogen´s more attractive core potential lowers the onset energies of valence and conduction band and turns the N-doped GNRs into n-type semiconductors. On the other hand, due to the particular GNR and substrate combination, undoped GNRs on Au are p-doped, displaying their valence band close to the Fermi level. Combination of pristine and N-doped segments into the same nanoribbon thus creates sharp p-n heterojunctions, with a band offset of around 0.5 eV for the heterojunction displayed in Fig. 15b. Because the band bending occurs over a distance in the order of 2 nm, the resulting electric field at the interface is extremely high ($2 \times 10^8$ V/m), making these heterostructures highly promising for electronic device based on p-n junctions [89].

Type I heterojunctions have been synthesized combining precursors that lead to differently wide GNRs [98]. The resulting structure is thus a width-modulated GNR as shown in Fig. 15a, with segments of 7-aGNRs and of 13-aGNRs. The latter has a larger band gap than the former, resulting in the straddling gap evolution that characterizes type I heterojunctions. Because the 7-aGNR segments serve as energy barrier for charge carriers localized along the 13-aGNR segment, the formation of quantum well states can be seen for short 13-aGNR segments surrounded by the narrower 7-aGNR, establishing an ideal framework for potential band gap engineering [97]. Furthermore, quantum dots could potentially be used as active components inside GNRs.

An alternative way for the creation of quantum dots embedded inside 7-aGNRs has been demonstrated by mixing pristine 7-aGNR with a small amount of boron doped 7-aGNRs precursors (Fig. 16a) [100]. The pristine regions in these hybrid ribbons preserve the electronic structure of 7-aGNR, while the borylated segments lack an energy level aligned with the pristine VB. As a result, the VB electrons on the pristine segments become confined by the boron atom pairs, since the VB ends abruptly over the borylated regions (Fig. 16b).

The calculated transmission function (Fig. 16c) for these free-standing hybrid 7-aGNR, that is, the transmission of electrons from one side of the ribbon to the other, shows that the boron pairs are very efficient reflectors and that the VB-1 acts as a transmission channel even though the VB is confined (Fig. 16d). This actually implies that boron atoms selectively confine VB electrons while leaving the VB-1 unaffected. The reason behind this selectivity stems from the symmetry between the boron induced states and the 7-aGNR bands [100]. These results highlight that the use of substitutional heteroatoms as dopants, and their related heterostructures, goes beyond a simple charge doping model and can strongly modify the transport properties of GNRs.

## 5 Applications

Graphene nanoribbons have been used in many types of applications as energy storage [101,102], thin films coatings [103-105] or in composite materials [106,107,108], but thanks to their attractive physical properties it is in electronic and optoelectronic devices applications where they arouse most attention. As a result, research efforts have pushed towards the integration of GNRs as active components in electronic devices like sensors [109,110], photodetectors [111-113] and field effect transistors (FETs) [114-119].

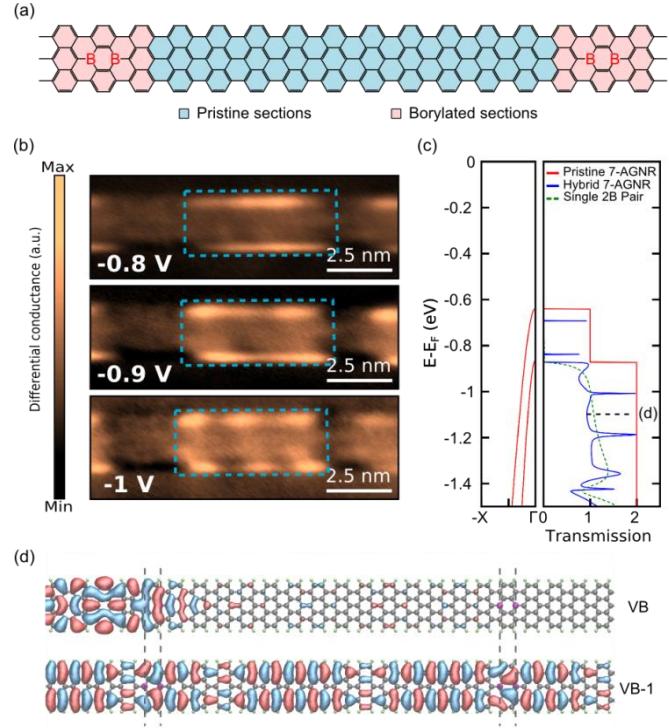

**Fig. 16 a** Schematic representation of the hybrid 7-aGNR, consisting in extended pristine regions and disperse borylated segments. **b** dI/dV maps of a pristine region (dashed blue square) enclosed between two borylated segments. The maps show an increasing number of modes in the conductance with increasing negative bias, fingerprint of the VB confinement. **c** Band structure of a pristine 7-aGNR (left) and transmission function (right) of a 7-aGNR (red) and a hybrid 7-aGNR (blue). The sharp peaks of transmission are related to the quantum well levels caused by the VB confinement. The step-like increase of transmission above the VB-1 onset reveals that the VB-1 is not confined. **d** Eigenchannel wave functions of the bands at the energy indicated in panel c. The VB electrons are strongly scattered, while the VB-1 electrons transmit almost freely. Adapted with permission from [100]. Copyright (2017) American Chemical Society.

A few experiments have already demonstrated the great potential of GNRs for technological applications. For example, nanometer-wide GNRs epitaxially grown on silicon carbide were found to be single-channel ballistic conductors at room-temperature over distances up to 16 micrometers, which is similar in performance to metallic carbon nanotubes (which are edgeless but for which the chirality determines the electronic structure) [120]. For thinner GNRs, the electron transport characteristics have been addressed locally in ultra-high-vacuum by STM with single ribbon precision [121]. Transport occurs in a tunneling regime and the tunneling decay length through a 7-aGNR measured at different bias voltages reveals the dependence of conductance with the ribbon electronic states, that is, energy level alignment and band gap value (Fig. 17a-b).

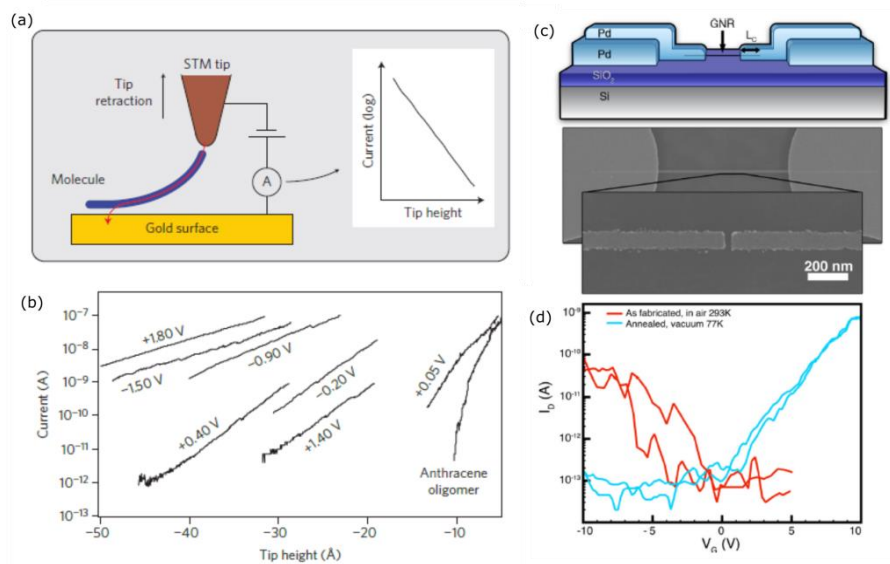

**Fig. 17 a** Schematic of the STM pulling experiment (arrow indicates tunnelling current). A characteristic current signal during the pulling sequence is shown in the right panel. **b** Current as a function of tip height for different experiments at different bias, evidencing the bias dependence of the conductance. **c** Schematic illustration of the device geometry and scanning electron micrograph of the device. **d** Electrical characterization of a typical device at $V_{SD} = 1$ V in both air and under vacuum at 77 K. (a-b) Reprinted by permission from Macmillan Publishers Ltd: Nature Nanotechnology [121], copyright (2012). (c.d) Reprinted from [118], with the permission of AIP Publishing.

Graphene nanoribbons technology is still in its infancy and a few problems need still to be solved to speed up the use of GNRs in current technology [4]. On-surface chemistry, as outlined in the previous pages, could overcome the problems of high quality fabrication and functionalization of GNRs. In fact, it allows producing graphene nanoribbons with atomically precise control of widths and edges, and thus with well-defined electronic structure. It allows to dope GNRs selectively and reproducibly, and eventually to increase their functionality by adding active components as photoactive, magnetic or switchable elements. For their operation in devices, as in gated multiterminal devices, GNRs have to be transferred to insulators. The direct growth on insulating substrates remains a challenge. Promising strategies may include the synthesis through photoactivation or through new reactions beyond the Ullman coupling - cyclodehydrogenation combination. Nevertheless, different routes can be explored to transfer GNRs from a metal to another material as silicon dioxide, as it has been successfully done for two-dimensional materials: exfoliation with or without a solution [118,119], intercalation assisted exfoliation [51], or chemical vapor deposition [122].

The last problem of GNR technology is the final fabrication and processing of the device. A few prototypes of field effect transistors have been realized exploiting on-surface synthesized GNRs (Fig. 17 c-d) [118,119]. The interest on FETs stems on the extreme thinness of GNRs which is expected to enable fabricating FETs with very short channels. This should significantly increase the device speed, while avoiding the unfavorable short-channel effects of standard CMOS technologies [123]. In GNR-FETs the driving current is limited by the contact resistance between a single GNR and the metallic contacts caused by the Schottky barrier at the interface. The height of the barrier depends linearly with the GNR band gap. In fact, FETs with 7-aGNR have shown low driving currents of 1 nA per 1 V drain bias [118] while FETs with GNRs displaying lower band gap like 9-aGNR and 13-aGNR exhibited on-currents up to 100 nA per -1 V bias and on/off ratios of $10^2$ to $10^5$ [119]. To further improve GNR-FETs performance wider ribbons with smaller band gaps could be used, or materials for metallic contacts could be tailored to the electronic structure of the particular GNR used. In addition, longer ribbons (lengths > 30 nm) would assure a better contact overlap with source and drain.

**Acknowledgments** Financial support by the European Research Council (ERC) under the European Union's Horizon 2020 research and innovation program (grant agreement No. 635919) and by the Spanish Ministry of Economy, Industry and Competitiveness (MINECO, Grant Nos. MAT2016-78293-C6-1-R, MAT2016-78293-C6-5-R and FIS 2015-62538-ERC) is acknowledged.